\documentclass[twoside,english]{article}
\usepackage[T1]{fontenc}
\usepackage[latin9]{inputenc}
\usepackage{geometry}
\usepackage{babel}
\usepackage{float}
\usepackage{url}
\usepackage{amsmath}
\usepackage{amssymb}
\usepackage{graphicx}
\usepackage{setspace}
\usepackage[authoryear]{natbib}
\usepackage[unicode=true,pdfusetitle,
 bookmarks=true,bookmarksnumbered=false,bookmarksopen=false,
 breaklinks=false,pdfborder={0 0 1},backref=false,colorlinks=false]
 {hyperref}

\makeatletter

\usepackage{babel}

\usepackage{babel}
\usepackage{url}
\usepackage{amsthm}

\numberwithin{equation}{section}
\numberwithin{figure}{section}

\usepackage{babel}
\usepackage{amsthm}
\usepackage{subcaption}

\numberwithin {equation}{section}
\numberwithin {figure}{section}

\AtBeginDocument{
  
}

\makeatother

\begin{document}
\title{Optimal transport on large networks \\
 a practitioner's guide}
\author{Arthur Charpentier\thanks{Mathematics Department, Universit� du Qu�bec � Montr�al (UQAM). Email:
charpentier.arthur@uqam.ca. Charpentier gratefully acknowledges funding
from NSERC grant 07077.} , Alfred Galichon\thanks{Economics Department, FAS, and Mathematics Department, Courant Institute,
New York University. Email: ag133@nyu.edu. Galichon gratefully acknowledges
funding from NSF grant DMS-1716489. Part of this research was carried
when he was visiting the Toulouse School of Economics.} , and Lucas Vernet\thanks{Economics Department, Sciences Po and Banque de France. Email: lucas.vernet@gmail.com.}}
\date{\today}

\maketitle
 
\begin{abstract}
This article presents a set of tools for the modeling of a spatial
allocation problem in a large geographic market and gives examples
of applications. In our settings, the market is described by a network
that maps the cost of travel between each pair of adjacent locations.
Two types of agents are located at the nodes of this network. The
buyers choose the most competitive sellers depending on their prices
and the cost to reach them. Their utility is assumed additive in both
these quantities. Each seller, taking as given other sellers prices,
sets her own price to have a demand equal to the one we observed.
We give a linear programming formulation for the equilibrium conditions.
After formally introducing our model we apply it on two examples:
prices offered by petrol stations and quality of services provided
by maternity wards. These examples illustrate the applicability of
our model to aggregate demand, rank prices and estimate cost structure
over the network. We insist on the possibility of applications to
large scale data sets using modern linear programming solvers such
as Gurobi. In addition to this paper we released a R toolbox to implement
our results and an online tutorial\footnote{\href{http://optimalnetwork.github.io}{http://optimalnetwork.github.io}}. 
\end{abstract}
\newpage{}

\section{Introduction}

The recent literature made an extensive use of networks to model market
structure. However the scalability of these models is limited by the
difficulty to compute equilibrium on large data sets. The results
are therefore dependent on possible edge effects. In this paper, we
present a set of tools to solve this issue and compute equilibrium
of spatial allocation problems in large geographic markets. In our
model we consider two types of agents, buyers and sellers, whose locations
are fixed. The market is modeled by a network that maps travel costs
between two adjacent locations. Buyers choose the most competitive
seller depending on their prices and the cost to reach them. Each
seller, taking as given other sellers prices, sets his own price to
have a demand equal to the one we observed. Under a linear programming
framework, the equilibrium is shown to be equivalent to a minimum
cost flow. This problem is a compact formulation of the matching problem
between buyers and sellers. After proving the existence of an equilibrium,
we discuss the question of uniqueness.

We illustrate the applicability of this model on two examples. In
the first one, knowing the prices set by gas stations and the distribution
of consumers over the market, we aggregate the demand at each station
and plot their geographic market share. The second example discusses
the competition between maternity wards over the quality of services
provided. Taking as given the location of households expecting a baby
and the number of births in each hospital, we infer a ranking of maternity
wards. For both of these examples we discuss our results and the limits
of our model. In addition to this paper we released a R toolbox and
an online tutorial\footnote{\href{http://optimalnetwork.github.io}{http://optimalnetwork.github.io}}
to simplify the application of these methods to economic problems.

\subsection*{Relation to literature}

Linear programming was pioneered by Kantorovich and Koopmans with
applications for the computation of economic optimum. In parallel,
Hitchcock formulated transportation problems as linear programs. Dantzig
developed the general framework for linear programming problems and
introduced the simplex method to solve them. We can note that from
the very beginning, both optimal transportation problems and optimum
of economic problems were the two important main applications of this
problem. This paper presents linear programming solvers that were
recently developed and can solve large scale problems. We defend the
idea that these tools renew the interest for a linear programming
formulation of economic problems.

Linear programming problems focus on the question of the economic
optimum whereas economic theory is more interested by economic equilibrium.
In matching markets with non-transferable \citep{gale1962college}
and transferable utility \citep{shapley1971assignment} equilibrium
is defined using the notion of pairwise stability. The duality theory
of linear programming shows that, when utility is transferable, the
solution of the optimization problem is an equilibrium.

Optimal transport problems were originally introduced by Monge and
solved by Kantorovich. More recently, \citet{bertsekas1998network}
studied optimal transport on networks and proposed several algorithms
to compute a solution. This theory is also developed in the literature
about the applications of optimal transport \citep{santambrogio2015optimal,galichon2016optimal}.
Several of these results started being used in recent developments
of the trade literature: \citet{eaton2002technology} developed quantitative
versions of the Armington trade models allowing counterfactuals with
respect to trade costs in a multicountry competitive equilibrium.
Following this work, \citet{allen2014trade}, \citet{allen2015existence}
and \citet{allen2016welfare} computed the first-order welfare impact
of reductions to the cost of shipping across specific links in an
Armington model but did not optimize over the space of networks. \citet{fajgelbaum2017optimal}
keep developing these models and make the connection with the optimal
transport theory, solving a global optimization over the space of
networks in a neoclassical framework. The dimensions of the problems
considered in these models remain however limited.

\section{Model}

\subsection*{Settings}

The geometry of the geographic market is described by a connected
graph $G=\left(V,E\right)$ with $V$ a set of vertices and $E$ a
set of directed edges. Vertices are the different locations; edges
are road segments linking two adjacent locations. Agents are located
at the vertices of our graph. We make the following assumptions on
their consumption preferences: 
\begin{description}
\item [{(A1)}] Consumers with the same location have the same outside option. 
\item [{(A2)}] Consumers have the same cost of travel along a given edge. 
\item [{(A3)}] There is no capacity constraints on the supply side. 
\end{description}
We will discuss in the next paragraph how these assumptions can be
relaxed. However they seem to be realistic on several markets and
in particular on both examples presented in this paper. They aim at
simplifying the computation of equilibria in large scale markets.

A path $\pi_{vw}$ between two nodes $v,w\in V$ is a sequence of
connected edges starting in $v$ and ending in $w$. We note $\Pi_{vw}$
the set of paths between $v$ and $w$. As $G$ is a connected graph,
the set of paths between any pair of nodes is not empty.

For each edge $e\in E$, we define $c_{e}$ the cost of travel along
this arc. A seller located in $w\in V$ sells at an endogenous price
$p_{w}$. Therefore a consumer located in $v$ chooses to buy from
the seller located in $w$ that minimizes his total cost $p_{v}$:
\[
p_{v}=\underset{w\in V\text{, }\pi_{vw}\in\Pi_{vw}}{\min}\left[p_{w}+\underset{e\in\pi_{vw}}{\sum}c_{e}\right]
\]

Let's denote $u_{v}$ the reservation utility of agents located in
$v$ and $p_{v}$ the price for these agents (including the cost of
transportation). Therefore, if $p_{v}>u_{v}$ a consumer in $v$ chooses
his outside option, else he buys from the most competitive seller
depending on his location in the market. This outside option can be
modeled by a fictitious node $0$, where the price of the commodity
is set to $p_{0}=0$ and the cost of travel is $c_{v0}=u_{v}$. Hence
the excess demand $z_{v}$ at a node $v\in V$ does not depend on
the price.

We are solving a matching problem over a network between sellers and
consumers. At equilibrium each consumer chooses the most competitive
seller depending on his location; furthermore the demand received
by each seller must equal his supply. If we define $\mu_{e}$ the
number of agents traveling along edge $e$, these equilibrium conditions
can be written: 
\[
\left\{ \begin{array}{l}
\forall vw\in E\text{, }p_{w}-p_{v}\leq c_{vw}\\
\forall vw\in E\text{ such that }\mu_{vw}>0\text{, }p_{w}-p_{v}=c_{vw}\\
\forall v\in V\text{, }\underset{w\in V}{\sum}\mu_{vw}-\underset{u\in V}{\sum}\mu_{uv}=z_{v}
\end{array}\right.
\]

These equations are the complementary slackness conditions for a minimum
cost flow problem over $G$. Hence, if the graph is connected and
if there isn't any loop of negative cost in $G$, there exists an
equilibrium which is the solution of the min-cost flow problem 
\[
\begin{array}{cl}
\mu=\arg & \min\underset{e\in E}{\sum}c_{e}\mu_{e}\\
 & \mu\text{ s.t. }\forall v\in V\text{, }\underset{w\in V}{\sum}\mu_{vw}-\underset{u\in V}{\sum}\mu_{uv}=z_{v}
\end{array}
\]

\subsection*{Degeneracy}

The optimal transport problem of Monge-Kantorovich is known for having
degenerated solutions. In economic terms this means that prices are
going to make consumers indifferent between several sellers. A power
diagram, also called Laguerre-Voronoi, is a partition of the space
that maps the geographic market shares of sellers. The non-uniqueness
of power diagrams is a consequence of the degeneracy of solutions
to the optimal transport problem. In this paragraph we explained how
the degeneracy can be mitigated at the price of some efficiency of
computation.

Consider for example the following market:

\begin{figure}[H]
\centering{}\includegraphics[scale=0.4]{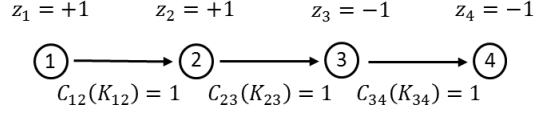}
\caption{Example of market with degenerated solutions}
\end{figure}
Buyers are located in nodes 1 and 2 and sellers in nodes 3 and 4.
At equilibrium $p_{4}=p_{3}+1$ and buyers are indifferent between
both sellers. Hence two degenerated power diagrams coexist as well
as any convex combination of both.

\begin{figure}[H]
\centering{}\includegraphics[scale=0.4]{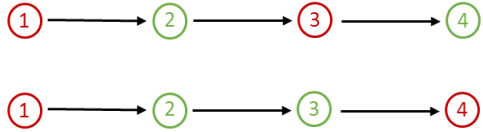}
\caption{Example of degenerated power diagrams}
\end{figure}
One way to mitigate this problem and select the relevant power diagrams
is to alter the cost of travel for agents. Let's consider that consumers
located is $v\in V$ solve the problem: 
\[
p_{v}=\underset{w\in V\text{, }\pi_{vw}\in\Pi_{vw}}{\min}\left[p_{w}+\left(\underset{e\in\pi_{vw}}{\sum}c_{e}\right)^{\alpha}\right]
\]
If $\alpha>1$ consumers value more the cost of closer arcs than the
ones that are further.

As the initial position of each agent matters, the equilibrium conditions
won't be the same as before. We now face a classical matching problem
with transferable utility. The computation of the equilibrium requires
first to calculate the geodesic distances between each pair of consumer
and seller. In the case of the example above, the problem becomes:

\begin{figure}[H]
\centering{}\includegraphics[scale=0.4]{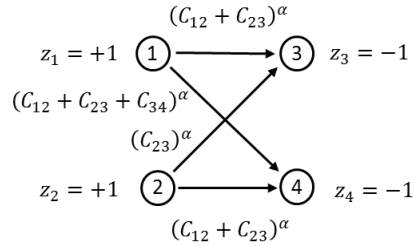}
\caption{Solving the degeneracy problem}
\end{figure}
Depending on the value of $\alpha$ we get a different power diagram
and solve the degeneracy problem at the price of increasing significantly
the dimensions of the problem. The network is less sparse than it
was and we have to solve a shortest past problems for each pair of
agents before solving the matching problem.

\begin{figure}[H]
\centering{}\includegraphics[scale=0.4]{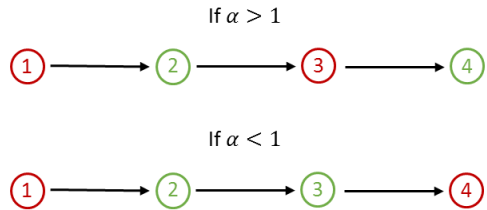}
\caption{Non-degenerated power diagrams}
\end{figure}

\subsection*{Build a geographic network}

The shapefile format is a normalized data format used by geographic
information system software. It spatially describes vector features
(points, lines and polygons) representing for example roads, lakes
or facilities. Each of these elements is associated with a vector
of characteristics that describes it. Several open source data sets
describing France geographic characteristics are distributed by the
IGN (Institut Geographic National) under a shapefile format. The examples
treated in this paper use the route500 dataset that describes 500
000 km of roads.

\begin{figure}[H]
\centering{}\includegraphics[scale=0.45]{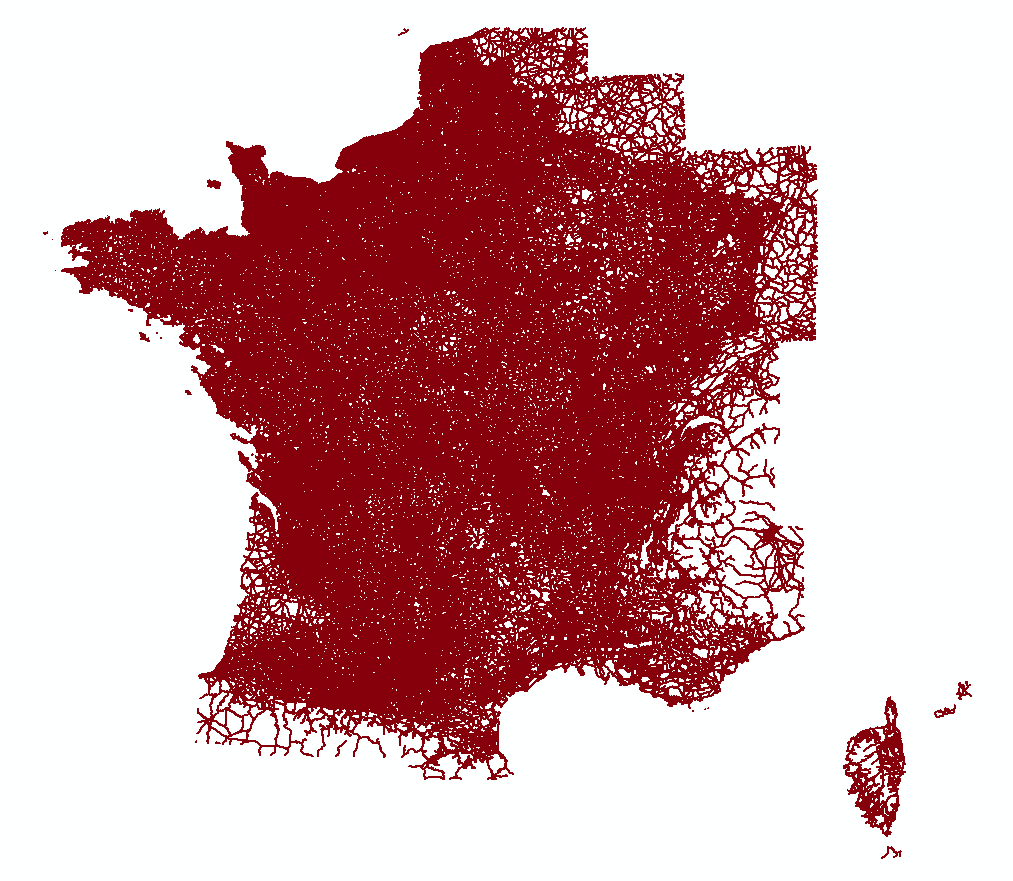} \caption{Example of shapefile: the route500 data set}
\end{figure}
We released a R toolbox that generates a network from a shapefile
and compute an equilibrium for the spatial matching problem described
above. Thus, the route500 dataset is converted in a directed graph
which contains 186 464 nodes and 583 550 arcs after simplification
and after retaining the largest connex subgraph. This graph maps the
french road network and the locations of our agents. Each node and
arc is associated with a vector of characteristics such as road type,
distance between extremities or administrative area. The following
figure shows the position of nodes within Paris.

\begin{figure}[H]
\centering{}\includegraphics[scale=0.28]{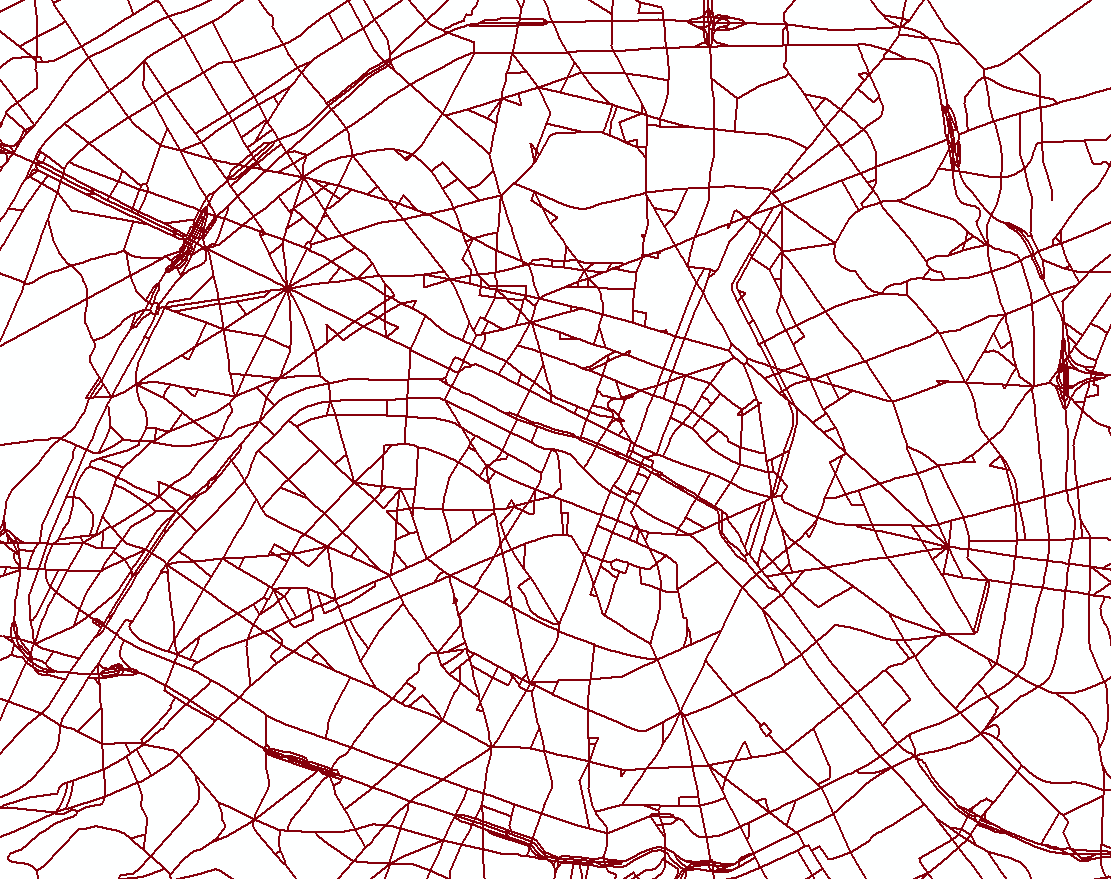}\quad{}\includegraphics[scale=0.28]{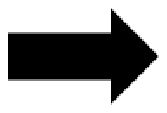}\quad{}\includegraphics[scale=0.33]{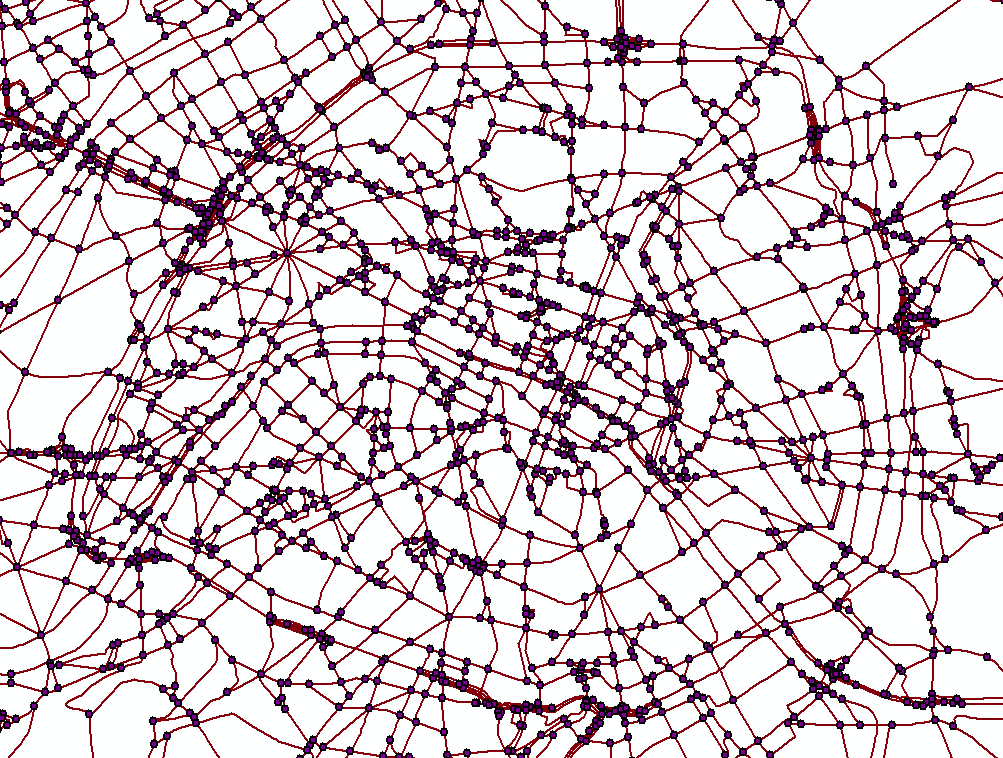}\caption{From a shapefile to a geographic network}
\end{figure}
The toolbox uses Gurobi, a particularly efficient linear programming
solver, to compute an equilibrium. For the network described above
and without parallel computing, it returns an equilibrium solution
in a few seconds. In addition to this toolbox, we also released a
comprehensive online tutorial\footnote{\href{http://optimalnetwork.github.io}{http://optimalnetwork.github.io}}
that goes through the examples presented in the last part of this
paper and describes the details of the computations.

\subsection*{Iceberg costs}

Models used in trade theory usually use iceberg transport costs. These
costs are linear in distance and paid in the transported good once
arrived - hence the metaphor of the iceberg that melts as you transport
it. Our model can incorporate these costs applying a log-transformation
to the equilibrium conditions.

Let's define for all $vw\in E$ iceberg costs $\tau_{vw}$. An equilibrium
$\left(\mu,p\right)$ should satisfy 
\[
\left\{ \begin{array}{l}
\forall vw\in E\text{, }\left(1-\tau_{vw}\right)p_{w}\leq p_{v}\\
\forall vw\in E\text{ such that }\mu_{vw}>0\text{, }\left(1-\tau_{vw}\right)p_{w}=p_{v}\\
\forall v\in V\text{, }\underset{w\in V}{\sum}\mu_{vw}-\underset{u\in V}{\sum}\mu_{uv}=z_{v}
\end{array}\right.
\]

which becomes 
\[
\left\{ \begin{array}{l}
\forall vw\in E\text{, }\log p_{w}-\log p_{v}\leq-\log\left(1-\tau_{vw}\right)\\
\forall vw\in E\text{ such that }\mu_{vw}>0\text{, }\log p_{w}-\log p_{v}=-\log\left(1-\tau_{vw}\right)\\
\forall v\in V\text{, }\underset{w\in V}{\sum}\mu_{vw}-\underset{u\in V}{\sum}\mu_{uv}=z_{v}
\end{array}\right.
\]

Therefore this problem is equivalent to the one presented in the first
paragraph of this section with $\forall vw\in E$, $c_{vw}=-\log\left(1-\tau_{vw}\right)$
and replacing $p$ by its logarithm.

\subsection*{Population heterogeneity}

The model can be altered to take into account some heterogeneity of
the agents. For example, if agents located a same node have different
reservation utility, this specific node can be replaced by two different
locations with the same connections to the rest of the network but
with different costs of transportation toward the node 0. Following
the same idea, we can model heterogeneity in the costs of transportation
across agents by duplicating some parts of the network.

\section{Applications to economic modeling}

As emphasize previously, the main interest of this model is its tractability
and our ability to compute solutions when its dimensions become large.
One difficulty we faced when we looked for examples to apply it, was
the difficulty to find disaggregated data. In this section, we focus
on two examples: prices offered by petrol stations and quality of
services provided by maternity wards. The first example is not new
in the economic literature and has been studied in \citet{borenstein1997gasoline}
and in \citet{eckert2004retail}.

\subsection*{Aggregate the demand}

Assume we know the cost of transportation, the locations of agents
and the prices set by sellers. To compute the demand served by each
seller we have to solve the discrete choice problem faced by consumers.
The result can be plotted on a power diagram that represents the geographic
market shares of each seller.

Let's consider the gas retail market in France. Gas stations, which
are on the supply side of this market, are assigned to the closest
node in the network built in the last section. The following figure
plots their locations and prices (the greener is a dot, the cheaper
is the price).

\begin{figure}[H]
\centering{}\includegraphics[scale=0.4]{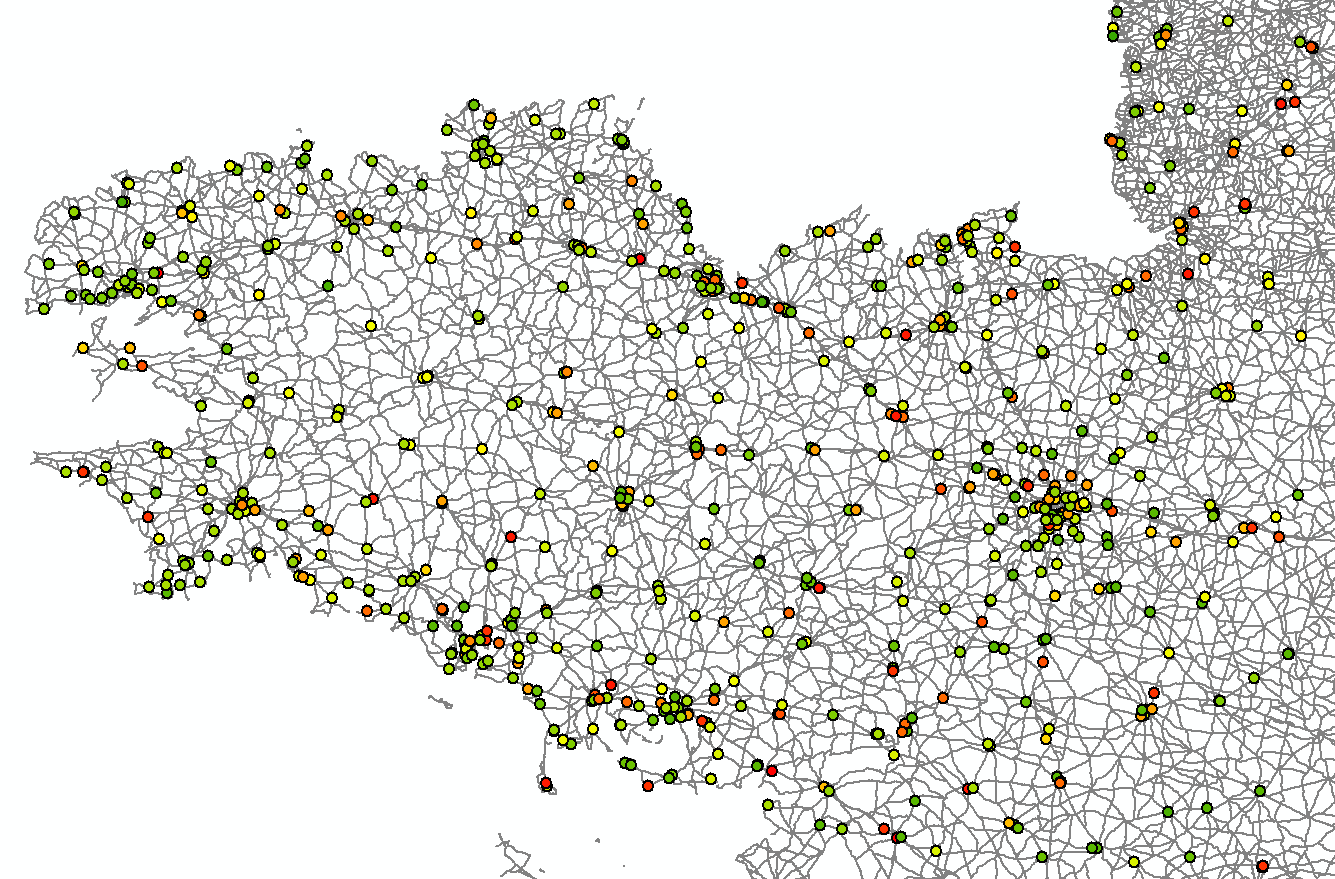}
\caption{Gas stations: location and price}
\end{figure}
Our proxy for the demand is the localized calls of the mobile application
\url{mon-essence.fr}. This application, installed on 1 to 5 millions
android mobiles, gives the prices in petrol stations that are close
to your location. We assume that each consumer wants to buy the same
quantity of fuel. This gives the following distribution of demand
within the market. Some criticisms can be made on this proxy and we
can in particular emphasize the risk of selection bias. As we don't
know the characteristics of the population that installed this application,
we can't control for this issue.

\begin{figure}[H]
\begin{centering}
\includegraphics[scale=0.25]{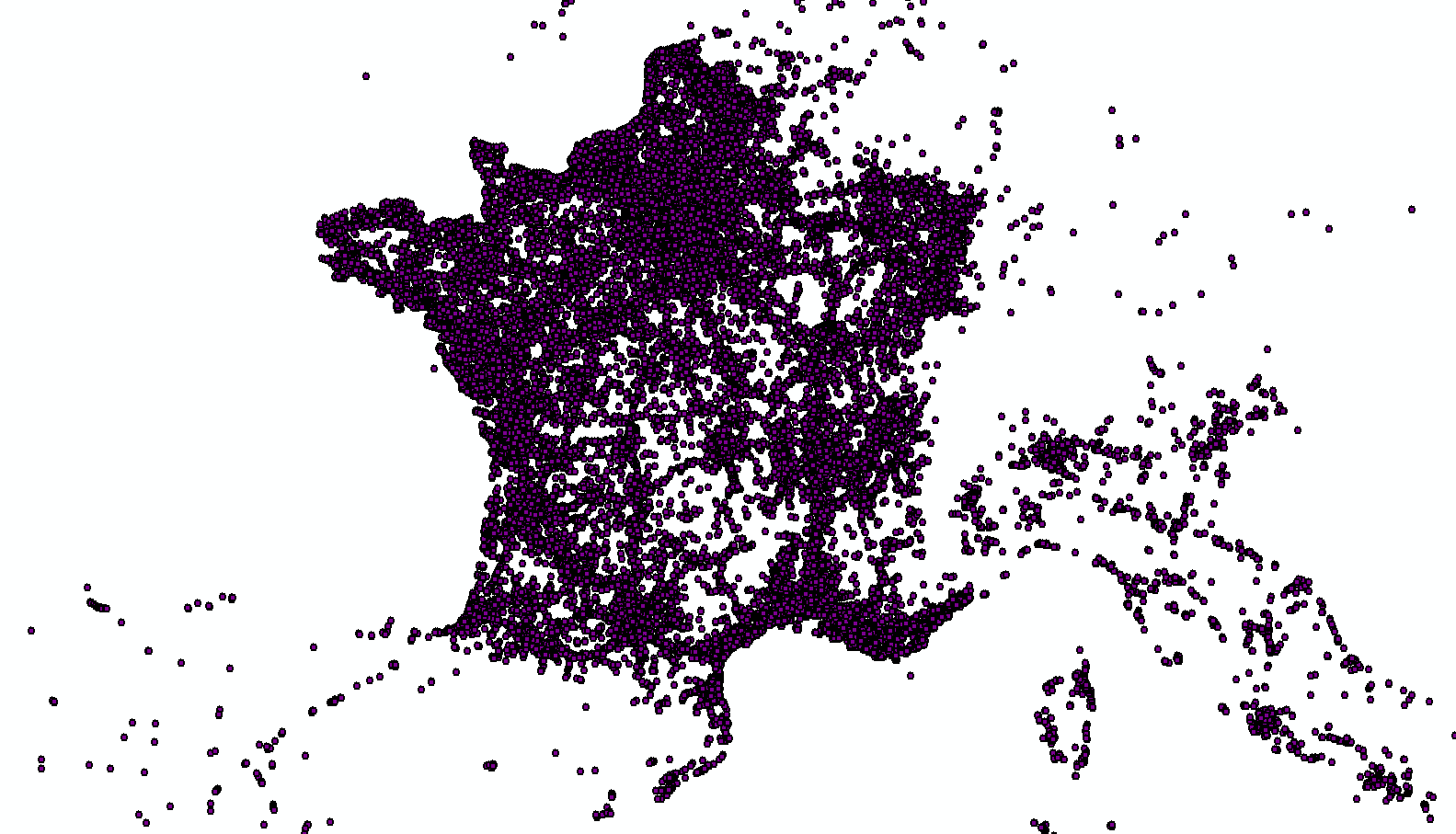} \includegraphics[scale=0.23]{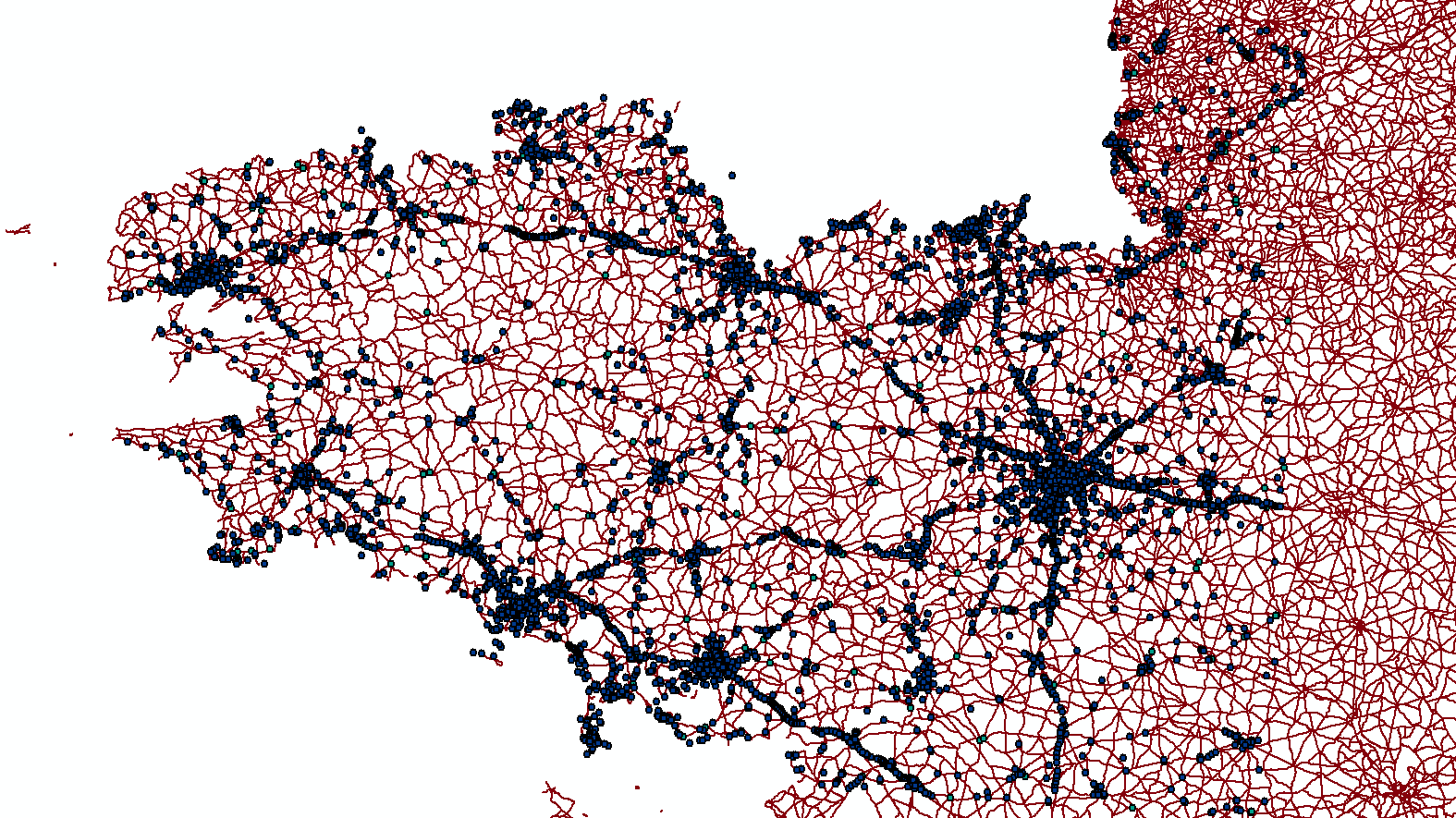} 
\par\end{centering}
\caption{Proxy for gas demand - France and Brittany}
\end{figure}
Finally, in order to compute the demand received by each gas station,
we need to assume a cost structure over the network. Here we multiply
the geographic distance between the two extremities of an arc by a
cost per kilometer. Obviously we could use other characteristics of
the edges such as the type of roads for example.

\begin{figure}[H]
\centering{}\includegraphics[scale=0.4]{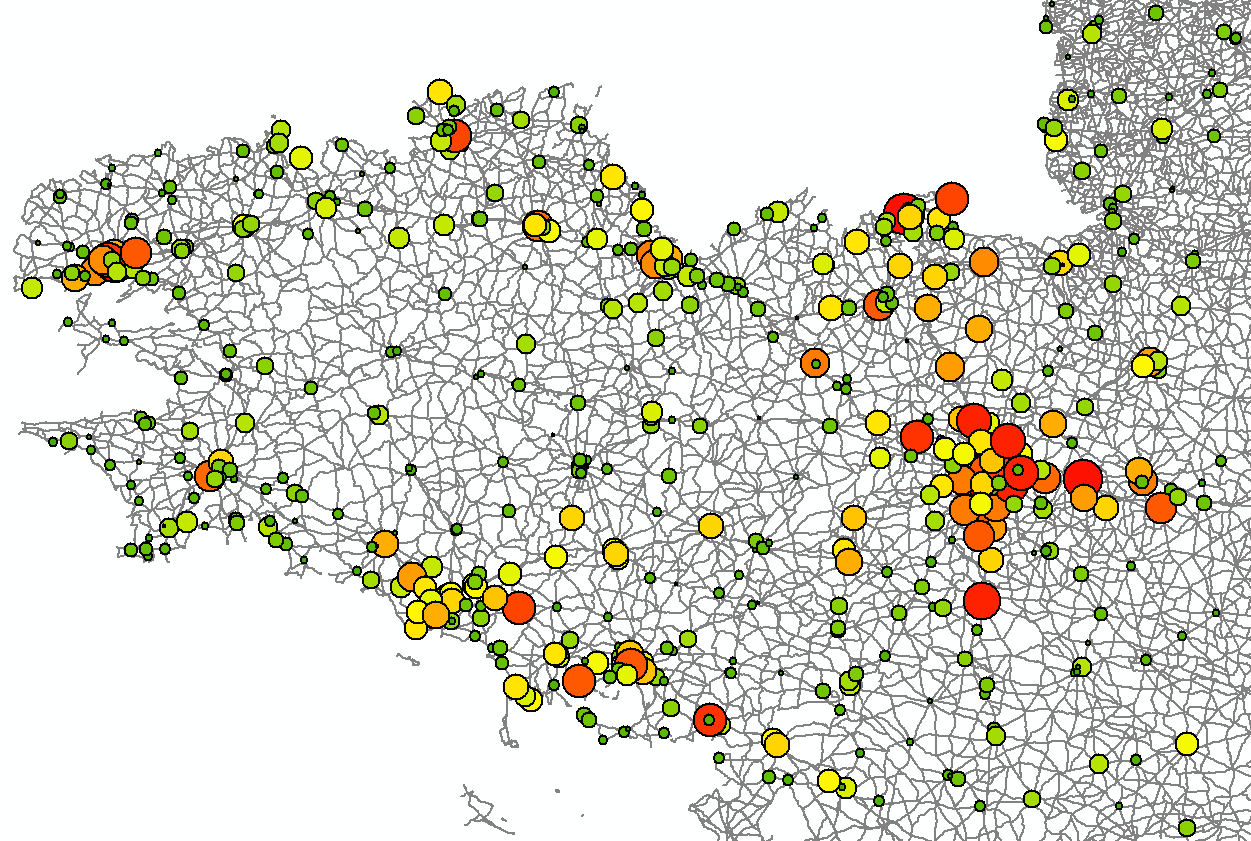}\caption{Demand received by each gas station}
\end{figure}
On smaller network, we can also plot the geographic market share of
each seller on a power diagram. The network structure may strongly
influence the shape and results we get.

\begin{figure}[H]
\centering{}\includegraphics[scale=0.19]{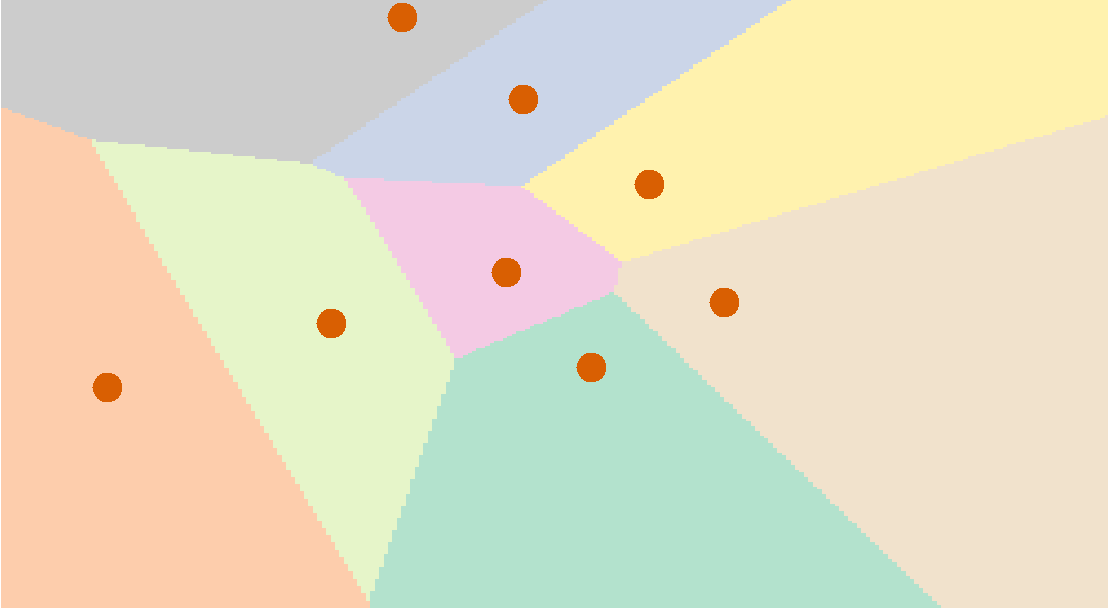}~~\includegraphics[scale=0.19]{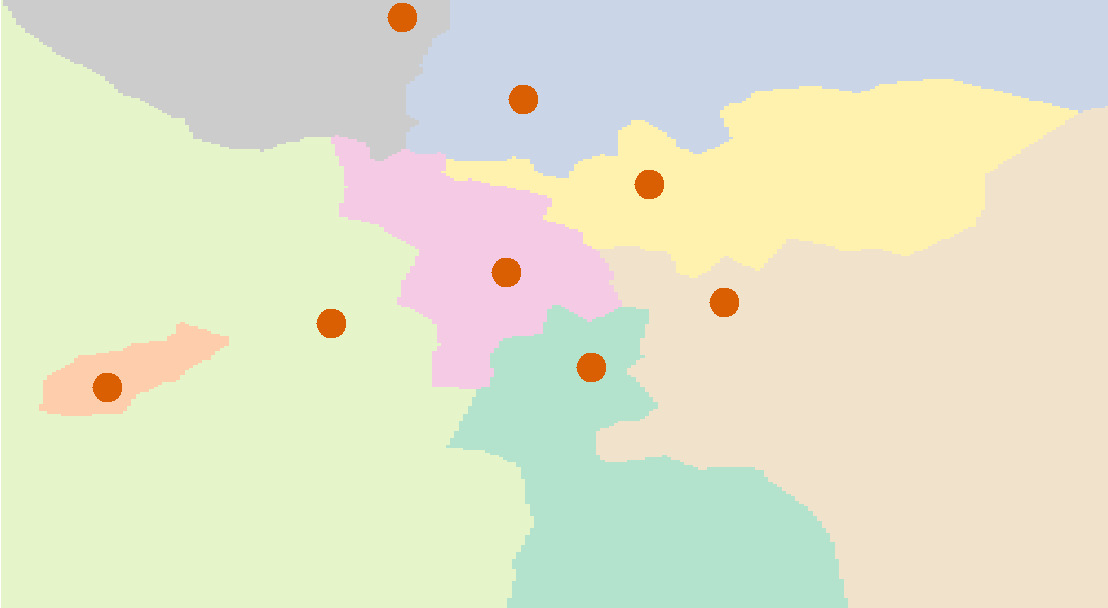}\hspace{0.5cm}
\includegraphics[scale=0.29]{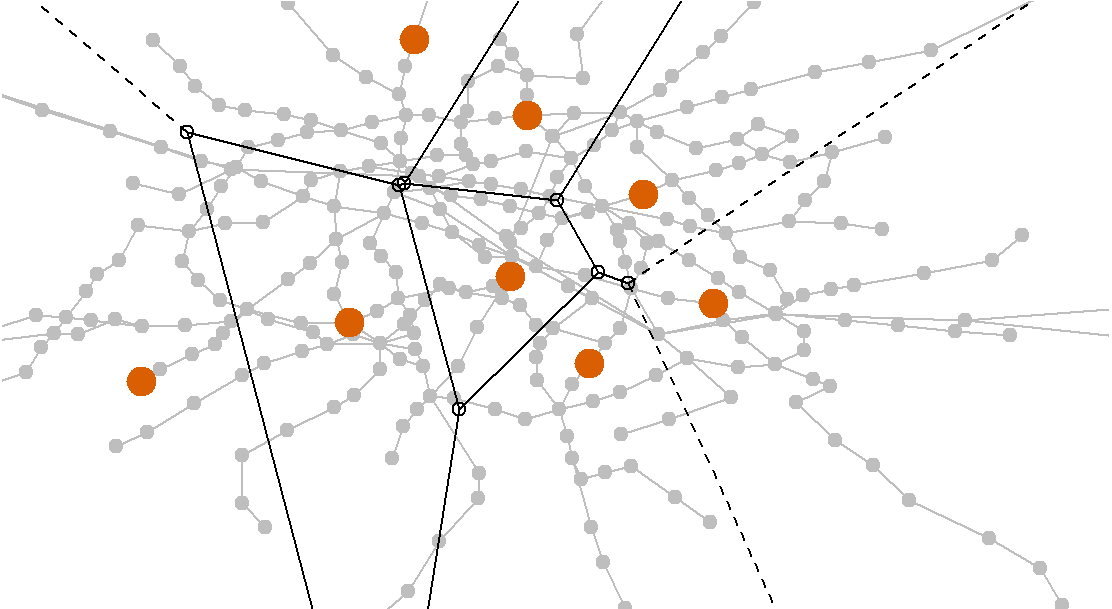}\vspace{0.25cm}
 \caption{Power diagrams with and without network}
\end{figure}
We can emphasize several limitations of the model. As said before,
this example was aiming at illustrating how one can find the demand
received by each seller knowing locations of consumers, prices and
travel costs. Here we make strong assumptions on the structure of
those three variables that should be more carefully discussed. The
fact that all consumers are willing to buy the same quantity of gasoline
is a particularly strong assumption. We should justify precisely the
use of our proxy for demand distribution and the travel costs structure.
Finally to avoid edge effects, we should not interpret results at
locations close to the borders of the network.

\subsection*{Rank the quality of service}

Health care costs are regulated in France and public hospitals can't
choose the price they charge. Patients choose their hospitals depending
on the travel cost and the quality of service provided. Hence the
prices we were considering in the last example are now the opposite
of an index for quality of service.

We know the location of pregnant households and the number of births
in each maternity ward and want to infer a ranking of hospitals. Once
again, we assume that travel cost along an edge is proportional with
its length. Because we only care about the order of maternity wards
ranked over their quality of service, the unit of the cost of transportation
does not matter.

We use the SAE data set (Statistiques Annuelles des Etablissements)
describing maternity wards in France. For each of them, we have its
location, the number of beds available, the number of workers in the
hospital, the number of births and the number of voluntary interruptions
of pregnancy per year. We cross this data set with the INSEE database
to get the number of mothers giving birth per administrative unit
between 2004 and 2014. We plot these data on the following figure.
On the left we draw the locations of maternity wards in France and
on the right the number of births per administrative unit in Paris,
in Ile-de-France and in Brittany.

\begin{figure}[H]
\begin{centering}
\begin{minipage}[c]{0.5\columnwidth}%
\begin{center}
\includegraphics[scale=0.45]{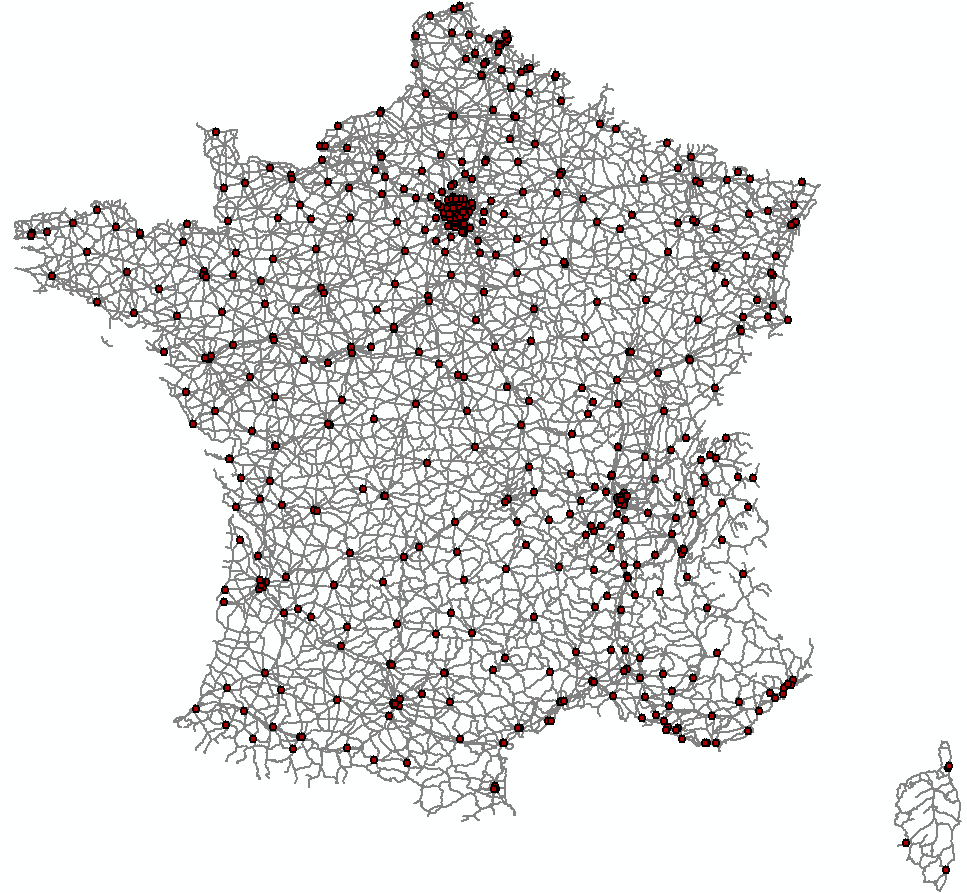} 
\par\end{center}%
\end{minipage}%
\begin{minipage}[c]{0.5\columnwidth}%
\begin{center}
\includegraphics[scale=0.23]{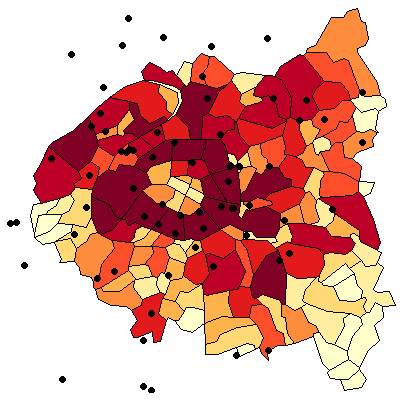}\includegraphics[scale=0.2]{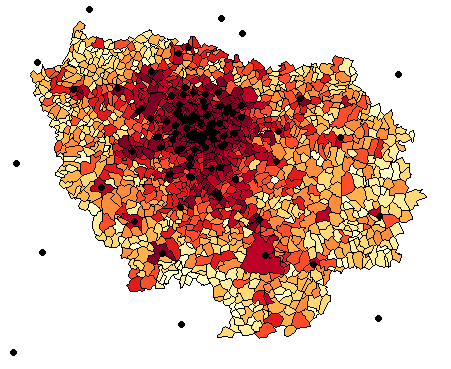}\\
 \includegraphics[scale=0.23]{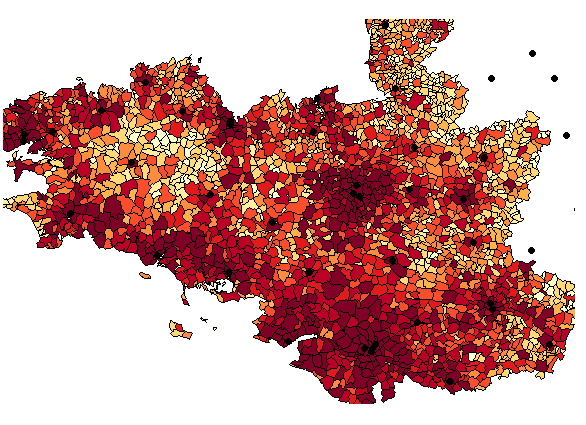} 
\par\end{center}%
\end{minipage}
\par\end{centering}
\caption{Location of maternity wards in France and number of births within
administrative units in Paris, in Ile-de-France and in Brittany}
\end{figure}
Within a given administrative area, we consider the number of roads
intersections as a reasonable proxy for the density of population.
Therefore, the population in an administrative unit is divided between
the different nodes inside this area.

One difficulty we met is the existence of outside options such as
private or foreign hospitals. We chose the arbitrary convention to
locally balance supply and demand - and therefore remove agents that
are using their outside options - over five geographic areas: North-West,
North-East, South-West, South-East and Paris area. We solve the associated
min-cost flow problem and plot the ranking of maternity wards on the
following figure.

\begin{figure}[H]
\centering{}\includegraphics[scale=0.6]{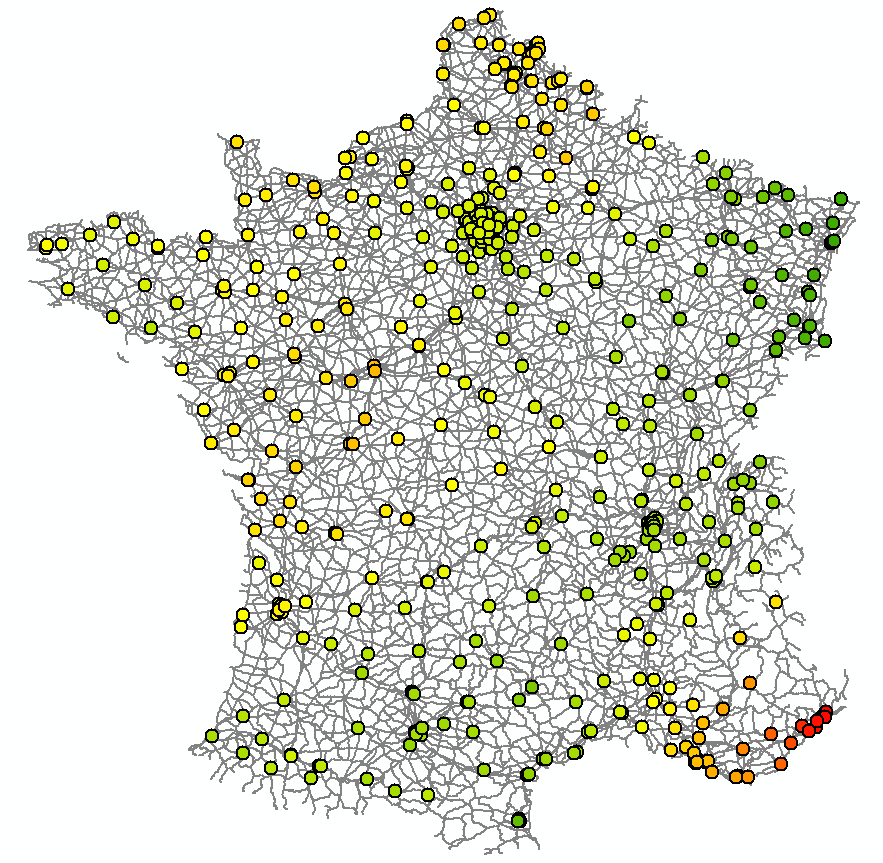}\caption{Quality of services in maternity wards}
\end{figure}
We see two main limitations to this model. Maternity wards may face
binding capacity constraints such as number of doctors or available
beds. We would also need additional data on the localization of outside
options such as private hospitals or information to remove agents
using their outside option from our data set.

\section{Conclusion}

In this paper we present a set of tools for the economic modeling
of matching problems between two types of agents located on a large
geographic market. We believe that the recent development of efficient
linear programming solvers such as Gurobi could renew the interest
of economists for these methods.

The modeling of a geographic market as a network seems relevant but
also practical and efficient when dimensions become large. This efficiency
is a consequence of the natural sparsity of roads network but also
a consequence of the strong assumptions of the model such as linear
costs of transport and additive utility. Whereas this assumptions
obviously brings limitations, we consider that the advantages of tractability
exceed them.

The examples presented in the paper are in line with this idea. The
difficulty to find disaggregated data on large geographic market make
us focus on two different examples: the gas retail market and the
maternity wards. We used these examples to present several applications
such as the computation of demand received by each gas retailer and
the ranking of the quality of service in maternity wards.

In addition to this paper we released a R toolbox to implement our
results and an online tutorial\footnote{{\href{http://optimalnetwork.github.io}{http://optimalnetwork.github.io}}}.

  \bibliographystyle{apalike}
\nocite{*}
\bibliography{Biblio/BiblioCompetitionAndPricingInNetworks}

\end{document}